\documentclass[fleqn,12pt,twoside]{article}
\usepackage{espcrc1}

\usepackage{graphicx}
\usepackage[figuresright]{rotating}

\newcommand{\AmS}{{\protect\the\textfont2
  A\kern-.1667em\lower.5ex\hbox{M}\kern-.125emS}}

\title{Extracting the $rp$-process from X-ray burst light curves}%
\author{J.L.~Fisker\address[ND]{Department of Physics and Joint Institute for Nuclear Astrophysics, University of Notre Dame, Notre Dame, IN 46556}\thanks{JLF is supported through the Joint Institute of Nuclear Astrophysics by NSF-PFC grant PHY02-16783.}, 
    E. Brown\address[MSU1]{Department of Physics and Astronomy and Joint Institute for Nuclear Astrophysics, Michigan State University, East Lansing, MI 48824-2320},
    M. Liebend\"orfer\address{CITA, University of Toronto, Ontario M5S 3H8, Canada},
    H. Schatz\address[MSU2]{Department of Physics and Astronomy and National Superconducting Cyclotron Laboratory and Joint Institute for Nuclear Astrophysics, Michigan State University, East Lansing, MI 48824-2320}\thanks{HS is an Alfred P. Sloan Fellow and is supported by NSF grants PHY 02-16783 (JINA) and PHY 01-10253 (NSCL).}
    and F.-K.~Thielemann\address[Basel]{Department of Physics and Astronomy, Klingelbergstrasse 82, 4056 Basel, Switzerland}\thanks{FKT is supported by Swiss NSF grant 20-068031.02.}}
       
\begin{document}
\bibliographystyle{h-elsevier}

\maketitle

\begin{abstract}
The light curves of type I X-ray bursts (XRBs) result from energy released from the atmosphere of a neutron star when accreted hydrogen and helium ignite and burn explosively via the $rp$-process.
Since charged particle reaction rates are both density and very temperature dependent, a simulation model must provide accurate values of these variables to predict the reaction flow.
This paper uses a self-consistent one-dimensional model calculation with a constant accretion rate of $\dot{M}=5\times 10^{16}\textrm{g}/\textrm{s}$ ($0.045 \dot{M}_{Ed.}$) and reports on the detailed $rp$-process reaction flow of a given burst.
\end{abstract}

\section{Introduction}
The $rp$-process in XRBs involves thousands of different reaction rates, which include $(p,\gamma)$-, $(\alpha,\gamma)$-, and $(\alpha,p)$-reactions, their respective inverse reactions, and electron captures and $\beta^+$-decays, and produces proton-rich and highly radioactive nuclei near or at the proton dripline \cite{Wallace81}. 
Most of these reactions have not been measured, but with recent upgrades of existing experimental facilities and the construction of new facilities such as the proposed RIA and GSI-FAIR more reactions are becoming accesible to experimentalists 
\cite{Wiescher01b}. 
Yet radioactive ion beam experiments take a long time to prepare and execute, so in order to prioritize experiments it is important to accurately establish the reaction flow to determine which reactions are the most critical in determining the x-ray burst.

With a neutron star surface gravity of $g \sim 10^{14}\textrm{cm}/\textrm{s}^2$ maintaining a hydrostatic pressure balance requires extreme densities and temperatures which are otherwise not obtainable in terrestrial laboratories.
Thus accreting neutron stars are also excellent nuclear laboratories, because the emitted X-ray light curve depends indirectly on the nuclear reactions in different layers of the neutron star atmosphere \cite{Wiescher02}.

The observed X-ray burst light curve obtains from the outbound heat transport of the nuclear burning front as it spreads around the star \cite{Lamb00}, 
whence direct comparisons require multi-dimensional modeling which is still computationally unaffordable. 
Therefore comparisons between models and observations are restricted to gross burst features such as rise and decay times for bolo-metrically single peaked XRB light curves.
Luckily the local nuclear reaction flow, which will be an important ingredient in next-generation computational models and a guide to future experiments, depends only on the local values of temperature, density and composition. 
Currently these values are best determined by self-consistent one-dimensional models.

In this paper we use such a model which is detailed in \cite{Fisker04}.
It includes a 298 isotope network, diffusive heat transport and convection, and solves the full general relativistic equations on a co-moving metric with a conservative discretization.
The computational domain is discretized into 103 zones with a pressure ranging from $P=5.1\times 10^{20}\textrm{erg}\cdot\textrm{cm}^{-3}$ to $P=6.7\times 10^{23}\textrm{erg}\cdot\textrm{cm}^{-3}$ and it is bounded by a relativistically corrected radiative zero atmosphere and a realistic core boundary interface.

\section{The nuclear reaction flow}
This paper considers a neutron star accreting at a constant rate of $\dot{M}=5\times 10^{16}\textrm{g}\textrm{s}^{-1}$.
At this accretion rate ignition occurs in a layer where hydrogen has burned completely into helium \cite{Bildsten98c}. 
Consequently the light curve obtains from the catastrophic detonation of a helium-rich layer and the subsequent conflagration of a layer of mixed H/He located above.
Depending on the violence of the helium detonation and the burning time-scale and the strength of the mixed H/He burning a double peaked burst can obtain \cite{Fisker04b}.

Ignition occurs in the ashes of the previous burst at the hydrogen fuel surface (see \cite{Fushiki87}), where the concentration of ${}^{4}\textrm{He}$ is maximal with mass fraction, $X_\alpha=0.57$. 
The rapidly released heat spreads inwards and outwards demarcating two burning regions: One comprising helium burning where no hydrogen is present and the other comprising hydrogen burning in the form of the $rp$-process.
\subsection{Helium flash}
Unlike the $rp$-process, helium burning does not depend on temperature-independent $\beta$-decays, whence the helium burning region rapidly adjusts within the tenths of a second it takes to reach peak temperature of $T=1.4\times 10^9\textrm{K}$. 
As the runaway exhausts ${}^4\textrm{He}$, the triple-alpha slows down markedly, because the helium burning triple-alpha process is very sensitive to the temperature and the ${}^4\textrm{He}$ density.
Other reactions include $(\alpha,p)$-reactions on previously generated stable ashes between ${}^{23}\textrm{Na}$ and ${}^{34}\textrm{S}$, where the released protons capture on the alpha-chain nuclei ${}^{12}\textrm{S}$, ${}^{16}\textrm{S}$, ${}^{20}\textrm{Ne}$, ${}^{24}\textrm{Mg}$, ${}^{28}\textrm{Si}$, ${}^{32}\textrm{S}$, and ${}^{36}\textrm{Ar}$ until the Coulomb barrier prohibits further $(\alpha,p)$-reactions on the resulting compound nuclei.

The rapid temperature rise obtained from the helium detonation establishes a slightly super-adiabatic temperature gradient which causes a rapid rise in surface luminosity and maintains a convective region for 0.9 seconds
This region extends all the way to the upper boundary of our computational domain and mixes its composition thoroughly.
Since the helium burns in about a tenth of a second, the subsequent energy release of the XRB obtains from the $rp$-process in the mixed layer of H/He above. 
\subsection{Mixed H/He burning}
Prior to the burst the composition of this region ranges from freshly accreted material to processed material of previous bursts which has been advected down from the surface while burning steadily in the hot CNO-cycle.
The heat of the convective bubble of the helium detonation rapidly mixes the entire region which consequently assumes the same composition. 
Therefore the end-product of the burning depends on the temperature attained in a given layer. 

The upcoming heat wave triggers the triple-alpha reaction and the ${}^{14}\textrm{O}(\alpha,p){}^{17}\textrm{F}$ and ${}^{15}\textrm{O}(\alpha,\gamma){}^{19}\textrm{Ne}$ breakout reactions. 
The former establishes the hot-CNO bi-cycle of \cite{Wallace81}, which breakout depends on the ${}^{18}\textrm{Ne}(\alpha,p)$ ${}^{21}\textrm{Na}$, whereas the latter establishes a flow to ${}^{21}\textrm{Mg}$ via ${}^{19}\textrm{Ne}(p,\gamma)$ ${}^{20}\textrm{Na}(p,\gamma)$ ${}^{21}\textrm{Mg}$, where it is blocked by photo-disintegration, because of the ${}^{21}\textrm{Mg}(p,\gamma)(\gamma,p){}^{22}\textrm{Al}$-equilibrium. 
Therefore the flow proceeds via ${}^{21}\textrm{Mg}(\beta^+, T_{1/2}=0.122\textrm{s})$ ${}^{21}\textrm{Na}(p,\gamma)$ ${}^{22}\textrm{Mg}$, where it branches going into either ${}^{22}\textrm{Mg}(\beta^+, T_{1/2}=3.86\textrm{s})$ ${}^{22}\textrm{Na}(p,\gamma)$ ${}^{23}\textrm{Mg}(p,\gamma)$ ${}^{24}\textrm{Al}$ or ${}^{22}\textrm{Mg}(p,\gamma)$ ${}^{23}\textrm{Al}(p,\gamma)$ ${}^{24}\textrm{Si}(\beta^+, T_{1/2}=0.102\textrm{s})$ ${}^{24}\textrm{Al}$. 
${}^{24}\textrm{Al}$ captures a proton to ${}^{25}\textrm{Si}$, which is again in blocked by photo-disintegration from the ${}^{25}\textrm{Si}(p,\gamma)(\gamma,p){}^{26}\textrm{P}$ reaction, where a fast additional proton capture may create the short-lived ${}^{27}\textrm{S}$, which decays to ${}^{27}\textrm{P}$. 
Following the ${}^{25}\textrm{Si}(\beta^+, T_{1/2}=0.634\textrm{s})$ decay, the reaction path proceeds via ${}^{25}\textrm{Al}(p,\gamma)$ ${}^{26}\textrm{Si}$ $(T_{1/2}=2.23\textrm{s})$, which is too long-lived for its decay to be relevant, so an additional proton capture leads to ${}^{27}\textrm{P}$.
Here the flow branches into either ${}^{27}\textrm{P}(\beta^+, T_{1/2}=0.260\textrm{s})$ ${}^{27}\textrm{Si}(p,\gamma)$ ${}^{28}\textrm{P}$ or ${}^{27}\textrm{P}(p,\gamma)$ ${}^{28}\textrm{S}(\beta^+, T_{1/2}=0.125\textrm{s})$ ${}^{28}\textrm{P}$.
This is followed by ${}^{28}\textrm{P}(p,\gamma)$ ${}^{29}\textrm{S}(\beta^+, T_{1/2}=0.187\textrm{s})$ ${}^{29}\textrm{P}(p,\gamma)$ ${}^{30}\textrm{S}(\beta^+, T_{1/2}=1.18\textrm{s})$ where both of the $(p,\gamma)$-reactions on the sulfur nuclei have low $Q$-values, whence the chlorine compound nuclei are immediately photo-disintegrated. 

The time-scale of this flow depends on the arithmetic sum of the $\beta^+$-decay half lives of the waiting points in the reaction path \cite{Wormer94}. 
These waiting points are circumventable by the the $(\alpha,p)$-process which makes the flow much faster \cite{Wallace81}. 
$(\alpha,p)$-reactions, which are very temperature-dependent, occur on the following isotopes: 
${}^{18}\textrm{Ne}$, ${}^{21}\textrm{Mg}$,${}^{22}\textrm{Mg}$, ${}^{24}\textrm{Si}$, ${}^{25}\textrm{Si}$, and ${}^{26}\textrm{Si}$.
At this point it becomes difficult for the alpha-particle to penetrate the Coulomb barrier of the target, but depending on the exact reaction rate of $(\alpha,p)$-reactions on ${}^{28}\textrm{S}$, ${}^{29}\textrm{S}$, ${}^{30}\textrm{S}$, ${}^{32}\textrm{Ar}$, ${}^{33}\textrm{Ar}$, and ${}^{34}\textrm{Ar}$ it may be possible to extend the flow further. 
Details of these reactions and their effect on the observable light curve are discussed in \cite{Fisker04b}.

Following the flow from ${}^{30}\textrm{S}$, it branches into either ${}^{30}\textrm{S}(\alpha,p)$ ${}^{33}\textrm{Cl}$ or ${}^{30}\textrm{S}(p,\gamma)(\gamma,p)$ ${}^{31}\textrm{Cl}(\beta^+, T_{1/2}=0.200\textrm{s})$ ${}^{31}\textrm{S}(p,\gamma)$ 
${}^{32}\textrm{Cl}(\beta^+, T_{1/2}=0.298\textrm{s})$ ${}^{32}\textrm{S}(p,\gamma)$ ${}^{33}\textrm{Cl}$ or 
${}^{30}\textrm{S}$ $(T_{1/2}=1.18\textrm{s})$ ${}^{30}\textrm{P}(p,\gamma)$ ${}^{31}\textrm{S}(p,\gamma)$ ${}^{32}\textrm{Cl}(\beta^+, T_{1/2}=0.298\textrm{s})$ ${}^{32}\textrm{S}(p,\gamma)$ ${}^{33}\textrm{Cl}$.
From this point the flow continues with
${}^{33}\textrm{Cl}(p,\gamma)$ 
${}^{34}\textrm{Ar}(\beta^+, T_{1/2}=0.844\textrm{s})$
${}^{34}\textrm{Cl}(p,\gamma)$ 
${}^{35}\textrm{Ar}(p,\gamma)$ 
${}^{36}\textrm{K}(\beta^+, T_{1/2}=0.342\textrm{s})$
${}^{36}\textrm{Ar}(p,\gamma)$ 
${}^{37}\textrm{K}(p,\gamma)$ 
${}^{38}\textrm{Ca}$. 
Following its $\beta$-decay and subsequent proton-capture the flow breaks into the $pf$-shell nuclei with ${}^{39}\textrm{Ca}(p,\gamma)(\gamma,p)$ ${}^{40}\textrm{Sc}(p,\gamma)$ ${}^{41}\textrm{Ti}(\beta^+, T_{1/2}=0.080\textrm{s})$ ${}^{41}\textrm{Sc}(p,\gamma)$ ${}^{42}\textrm{Ti}$.
Here the flow will branch again with either ${}^{42}\textrm{Ti}(\beta^+, T_{1/2}=0.199\textrm{s})$ ${}^{42}\textrm{Sc}(p,\gamma)$ ${}^{43}\textrm{Ti}(p,\gamma)$ ${}^{44}\textrm{V}$ or ${}^{42}\textrm{Ti}(p,\gamma)(\gamma,p)$ ${}^{43}\textrm{V}(p,\alpha)$ ${}^{44}\textrm{Cr}(\beta^+, T_{1/2}=0.053\textrm{s})$ ${}^{44}\textrm{V}$, where they connect and continue with ${}^{44}\textrm{V}(p,\gamma)$ ${}^{45}\textrm{Cr}(\beta^+, T_{1/2}=0.050\textrm{s})$ ${}^{45}\textrm{V}(p,\gamma)$ ${}^{46}\textrm{Cr}$, which decays to ${}^{46}\textrm{Ti}$ via ${}^{46}\textrm{V}$.

At the lower temperatures near the surface the $rp$-process effectively stops here, but at high temperature, the flow moves on via proton captures on these three $A=46$ isotopes, which comprise the first bottlenecks to the heavier iron-group nuclei.
The next bottleneck is the branching point ${}^{49}\textrm{Mn}$ which may either $\beta$-decay or proton capture. 
In the latter case the reaction flow out of ${}^{50}\textrm{Fe}$ is limited by photo-disintegration, so the main flow proceeds with two proton-captures on ${}^{51}\textrm{Fe}$ and ends up on ${}^{53}\textrm{Ni}$. 
The $Z=28$ Ni-isotopes are well-bound, whence the proton-capture $Q-$values are low. 
Depending on the $Q$-values, the flow can proceed via 
${}^{53}\textrm{Ni}(\beta^+, T_{1/2}=0.045\textrm{s})$ ${}^{53}\textrm{Co}(p,\gamma)$ ${}^{54}\textrm{Ni}(\beta^+, T_{1/2}=0.140\textrm{s})$ ${}^{54}\textrm{Co}(p,\gamma)$ ${}^{55}\textrm{Ni}(p,\gamma)(\gamma,p)$ ${}^{56}\textrm{Cu}(p,\gamma)$ ${}^{57}\textrm{Zn}$ otherwise the flow is stopped due to the long half-life of ${}^{56}\textrm{Ni}$.
An equivalent issue arises when the flow reaches ${}^{58}\textrm{Zn}$, ${}^{59}\textrm{Zn}$, and ${}^{60}\textrm{Zn}$ following ${}^{56}\textrm{Ni}(p,\gamma)$ ${}^{57}\textrm{Cu}(p,\gamma)$ ${}^{58}\textrm{Zn}$ and consecutive proton captures and $\beta$-decays.
The low proton-capture Q-values on ${}^{63}\textrm{Ge}$, ${}^{64}\textrm{Ge}$, ${}^{67}\textrm{Se}$, ${}^{68}\textrm{Se}$, ${}^{72}\textrm{Kr}$, and ${}^{76}\textrm{Sr}$ cause immediate photo-disintegration of the compound nuclei, so the flow and the resulting energy generation and burst decay luminosity is regulated by their comparably slow $\beta$-decays.
The maximum temperature is limited by the self-consistently determined hydrostatic pressure of the burning region and compared to the half-lives of the waiting points, the heat is quickly transported to cooler adjacent regions.
Therefore the flow does not extend further except for trace amounts leaving an average composition of $A\sim 63$ and $Z\sim 31$ thus corroborating the results of \cite{Woosley04}. 
\section{Conclusion}
Matching observational luminosity curves to computational models require detailed multi-D models, which are still computationally unavailable. 
However, gross luminosity features and the nuclear physics of the burst can be determined from 1D models. 
The burst rise time is determined by the characteristic time-scales of the burning regions. 
For mixed H/He burning, this is influenced by experimentally unknown reactions such as ${}^{30}\textrm{S}(\alpha,p){}^{33}\textrm{Cl}$ and ${}^{34}\textrm{Ar}(\alpha,p){}^{37}\textrm{K}$.
The surface composition following the burst depends on the proton-capture bottlenecks at $A=46$ and ${}^{49}\textrm{Mn}$. 
Finally the energy generation and composition of the burst ashes depend on the decays of the $A\geq 64$ waiting points whose distribution depends on the nuclear masses in that region. 


\end{document}